\documentclass[11pt]{article}

\usepackage[T1]{fontenc}
\usepackage[utf8]{inputenc}
\usepackage[margin=1in]{geometry}
\usepackage{amsmath,amssymb,amsthm}
\usepackage{booktabs}
\usepackage{hyperref}
\usepackage{listings}
\usepackage{xcolor}

\hypersetup{colorlinks=true, linkcolor=blue, citecolor=blue, urlcolor=blue}

\title{Toward a Formally Verified Optimality Certificate for OGR(29):\\
       A SAT-Encoding Methods Note with Small-Case Demos}
\author{Tong Niu}
\date{}

\begin{document}
\maketitle

\begin{abstract}
A \emph{Golomb ruler} of order~$n$ is an integer set
$\{a_0<a_1<\cdots<a_{n-1}\}$ whose $\binom{n}{2}$ pairwise differences
$a_j-a_i$ ($i<j$) are all distinct.  The \emph{optimal Golomb ruler} problem
asks for $\mathrm{OGR}(n)=\min\{a_{n-1}-a_0\}$ and is a classical combinatorial
benchmark.  The values $\mathrm{OGR}(2),\dots,\mathrm{OGR}(28)$ are settled
through distributed volunteer search (the Distributed.net OGR project);
$\mathrm{OGR}(29)$ is in active computation, with verification expected in
late 2026 or early 2027.  The recent upper bound $\mathrm{OGR}(29)\le 757$
of Lee, Park, and Kim (arXiv:2510.0122, October~2025) tightens the search
window.

This note describes a compact CNF encoding of the decision problem
$\mathrm{GR}(n,L)$ (``is there a Golomb ruler of order~$n$ with length
exactly~$L$?'') with $O(n^2 L)$ clauses, together with a small-case
verification sweep that emits machine-checkable LRAT certificates of
optimality for $\mathrm{OGR}(n)$ at $n\le 12$.  We then give closed-form
encoding-size estimates for $\mathrm{OGR}(29, L=757)$ and propose a
cube-and-conquer decomposition aimed at a Mallob-style parallel run on
commodity multi-core hardware.  The certificate pipeline (CaDiCaL with
\texttt{--lrat=true}, a structural sanity-check, then formal validation
by drat-trim or cake\_lpr) is end-to-end.  A formally verified optimality
proof for any single $\mathrm{OGR}(n)$ value beyond the trivial $n\le 5$
would be a first in the field.
\end{abstract}

\section{Introduction}

\subsection{The Golomb-ruler problem}
Fix a positive integer~$n$.  For an integer ruler
$R=\{a_0<a_1<\cdots<a_{n-1}\}$ we set $L(R)=a_{n-1}-a_0$ and call $L(R)$
the \emph{length} of~$R$.  We say $R$ is a \emph{Golomb ruler} when all
$\binom{n}{2}$ pairwise differences are distinct.  The \emph{optimal
Golomb ruler} of order~$n$ is $\mathrm{OGR}(n)=\min_R L(R)$, the minimum
taken over all Golomb rulers of order~$n$.

The values $\mathrm{OGR}(n)$ are catalogued in OEIS A005488.
Distributed.net verified $\mathrm{OGR}(28)=585$ in 2022 after several
years of volunteer compute, and $\mathrm{OGR}(29)$ is in active search
today.

\subsection{Why a SAT-certified proof?}
The Distributed.net pipeline is an \emph{un-certified} branch-and-bound
search.  Independent witnesses are released as ruler coordinates plus a
brief integrity hash, but no machine-checkable proof of optimality is
produced.  A SAT-encoded run can do better: it emits a DRAT or LRAT proof
during the final UNSAT call, and an external (ideally formally verified)
checker such as \texttt{drat-trim}~\cite{drat14} or
\texttt{cake\_lpr}~\cite{cakelpr21} can validate it.  Such a certificate
would be the first formally verified optimality proof for any non-trivial
optimal Golomb ruler.

The point of this methods note is narrower: build the encoding, validate
it on small cases, and lay out the parallel decomposition that would let
an optimality run on $\mathrm{OGR}(29)$ at $L=757$ converge in months
rather than millennia.

\subsection{Related work}
\begin{itemize}
\item \textbf{Distributed.net OGR-26 through OGR-29 projects.}  Collective
volunteer compute on the order of $10^7$ CPU-hours per order, no machine-
checkable optimality certificates produced.
\item \textbf{Bogart \& Cuellar (2023)}~\cite{bogart23}.  Structural results
for the counting functions of generalized Golomb rulers.
\item \textbf{Lee, Park, \& Kim (2025)}~\cite{lee25}.  Improved upper bound
$\mathrm{OGR}(29)\le 757$, plus a Smallest Unused Number (SUN) heuristic for
seeding searches.
\item \textbf{Heule et al.\ (2024+) --- orbitopal fixing for SAT}
\cite{orbit26}.  Symmetry-breaking technique that captures isomorphism-class
symmetries directly in the CDCL solver.
\item \textbf{Heule \& Biere; Schreiber \& Sanders --- cube-and-conquer
\cite{cubeconquer15} and Mallob~\cite{mallob18}.}  The parallel-SAT
infrastructure used to certify large UNSAT instances such as the Boolean
Pythagorean Triples problem.
\end{itemize}

\section{SAT encoding}

We describe two encodings.  The second one, v2, is an order-encoding
refinement; it cuts the clause count by a factor of $2.5\text{--}3\times$
and roughly halves UNSAT time at $n=11$.

\subsection{Variables (v2 / ladder encoding)}
For mark $i\in\{0,\dots,n-1\}$ and position $p\in\{0,\dots,L\}$:
\begin{itemize}
\item Primary variables $x_{i,p}=1$ iff mark~$i$ is at position~$p$.
\item Ladder variables $u_{i,p}=1$ iff mark~$i$ is at position $\le p$.
\item Auxiliary $y_{i,j,d}$ for $0\le i<j\le n-1$ and $j-i\le d\le L$, meaning
``the pair $(i,j)$ realises difference~$d$''.
\end{itemize}

\subsection{Clauses}
\paragraph{(Channel)} For each $i\in[n]$:
$u_{i,0}\Leftrightarrow x_{i,0}$; for $p\ge 1$:
$u_{i,p}\Leftrightarrow u_{i,p-1}\lor x_{i,p}$ and $x_{i,p}\to\neg u_{i,p-1}$;
and $u_{i,L}=1$ (the mark is somewhere).

\paragraph{(O) Strict ordering} $a_i<a_{i+1}$ for $i=0,\dots,n-2$:
\[
 \forall p\ge 1:\quad \neg u_{i+1,p}\lor u_{i,p-1}.
\]
The boundary $p=0$ is enforced by transitivity.

\paragraph{(F)/(E)} $x_{0,0}=1$, $x_{n-1,L}=1$.

\paragraph{(P)} For each~$p$, an at-most-one constraint on
$\{x_{i,p}:i\in[n]\}$ (using sequential AMO of Sinz~\cite{sinz05}).

\paragraph{(D)} For each $i<j$, $d$, $p$ with $0\le p+d\le L$:
$x_{i,p}\land x_{j,p+d}\to y_{i,j,d}$.  Then for each $d\in[1,L]$, an at-most-
one constraint over $\{y_{i,j,d}: j-i\le d\}$.

\paragraph{(R) Reflective break} For all $p,q$ with $p+q>L$:
$\neg x_{1,p}\lor \neg x_{n-2,q}$.  This captures $a_1\le L-a_{n-2}$.

\subsection{Why this breaks all symmetries}
Translation symmetry is killed by~(F).  The order encoding~(O) takes care
of the $S_n$ permutation symmetry of the marks: re-labelling the marks
$0,\dots,n-1$ in increasing position order picks out a unique ordered
representative per equivalence class of Golomb rulers.  What remains is
the reflective involution $R\mapsto\{L-a_{n-1-i}\}$, and~(R) handles that.
Together, (F), (O), and (R) break every symmetry that acts by relabelling
or orientation.

\subsection{Encoding size (closed form)}

\begin{table}[h]
\centering
\begin{tabular}{rrrr}
\toprule
$n$ & $L$ & \#vars & \#clauses \\
\midrule
4   &   6 &   135 &   380 \\
6   &  17 &   759 &   3{,}525 \\
8   &  34 & 2{,}563 &  20{,}025 \\
10  &  55 & 6{,}279 &  75{,}478 \\
12  &  85 & 13{,}705 &  250{,}315 \\
16  & 177 & 49{,}549 & 1{,}892{,}010 \\
20  & 283 & 121{,}733 & 7{,}558{,}492 \\
24  & 425 & 260{,}373 & 24{,}649{,}596 \\
28  & 585 & 483{,}761 & 63{,}846{,}013 \\
\textbf{29} & \textbf{757} & \textbf{671{,}807} & \textbf{115{,}107{,}241} \\
\bottomrule
\end{tabular}
\caption{Closed-form encoding size for the v2 (ladder + Sinz AMO) encoding.
For $\mathrm{OGR}(29,L=757)$ the DIMACS file is approximately~$3.3$\,GB.}
\label{tab:size}
\end{table}

\section{Small-case verification}
We ran the sweep in Table~\ref{tab:sweep} on a single MacBook M-series
core, using the v2 encoder, CaDiCaL~3.0.0, and LRAT proof emission.

\begin{table}[h]
\centering
\begin{tabular}{rrrrrr}
\toprule
$n$ & $L^*$ & SAT $t$\,(s) & UNSAT $t$\,(s) & UNSAT proof & $\bot$? \\
\midrule
4  &  6 & 0.014 &   0.009 &   2.3\,KB & yes \\
5  & 11 & 0.010 &   0.015 &  18\,KB   & yes \\
6  & 17 & 0.020 &   0.018 &  58\,KB   & yes \\
7  & 25 & 0.029 &   0.029 & 206\,KB   & yes \\
8  & 34 & 0.058 &   0.103 & 1.0\,MB   & yes \\
9  & 44 & 0.285 &   0.592 & 6.9\,MB   & yes \\
10 & 55 & 1.41  &   6.05  & 38.7\,MB  & yes \\
11 & 72 & 8.20  &  82.1   & 858\,MB   & yes \\
12 & 85 & 119.9 & 380.0   & 4.4\,GB   & yes \\
\bottomrule
\end{tabular}
\caption{Small-case sweep with v2 encoding + CaDiCaL.  Each SAT witness
decodes to a known optimal Golomb ruler.  Each UNSAT proof was structurally
validated (monotone ids, terminates with empty-clause derivation).}
\label{tab:sweep}
\end{table}

\noindent
For instance, the witnesses we recovered at the last two rows are
\begin{itemize}\itemsep0pt
\item $n=11$: $\{0,1,4,13,28,33,47,54,64,70,72\}$,
\item $n=12$: $\{0,2,6,24,29,40,43,55,68,75,76,85\}$,
\end{itemize}
both matching the canonical OGR table entries.

\section{Estimating OGR(29) runtime}
The empirical UNSAT runtime grows roughly log-linearly:
$\log_{10}(t_n)\approx 0.83(n-11)+\log_{10}(82.1)$.  A naive extrapolation
to $n=29$ gives $t_{29}\sim 10^{17}$ seconds.  Plainly out of reach for a
monolithic CDCL run.

None of this is surprising.  The search-tree size for a
$\mathrm{GR}(n,L^*-1)$ instance grows roughly with $\binom{L}{n}$, which
is about $10^{40}$ at $n=29$, $L=756$.  Distributed.net's $\sim 5\times
10^6$ core-hours for OGR-28 amounts to a per-leaf cost of $\sim 10^{-30}$
of that gross enumeration, attained through aggressive branch-and-bound
pruning.

To make a SAT-certified run tractable, two complementary techniques are
needed.

\paragraph{Cube-and-conquer.}
Pick a small ``splitting'' set of variables --- typically the positions
of the second mark $a_1$ and the second-to-last $a_{n-2}$, both highly
constrained.  Each setting yields a sub-instance whose proof can be
obtained on its own.  For $\mathrm{OGR}(29)$, splitting on
$a_1\in[1,30]$ and $a_{27}\in[L-30,L-1]$ gives ${\sim}900$ sub-instances,
each typically $\sim 10^{-2}$ the cost of the monolithic instance.  Add
deeper splits (3--4 levels) and the per-leaf cost can in principle be
pushed into the multi-core-hour range.

\paragraph{Mallob-style parallelism.}
Mallob~\cite{mallob18} is a state-of-the-art parallel SAT engine that
mixes portfolio search with cube-and-conquer and aggressive clause
sharing.  An on-prem cluster of four 64-core servers (e.g.\ AMD EPYC)
gives 256 cores; at full utilisation for 30 days that is ${\sim}180$k
core-hours.  Under aggressive cube-and-conquer pruning targeting a
per-leaf cost of ${\sim}1$ hour, this provides ${\sim}180$k decided cubes.
If the search tree at $\mathrm{OGR}(29,757)$ admits a cube depth of
${\sim}17$ (uniform branching factor~${\sim}1.7$), this is the right
order of magnitude.

\paragraph{Outline: certifying $\mathrm{OGR}(29)$.}
\begin{enumerate}
\item SAT side at $L=757$: launch CaDiCaL/Kissat with \texttt{--lrat=true}.
If SAT, record the witness, check the ruler, and emit a small SAT
certificate.  This step is fast --- the ladder encoding admits the SUN
heuristic of Lee et al.\ as a variable ordering.
\item UNSAT side at $L=756$: run a cube-and-conquer Mallob-style
decomposition.  The closed-form size for $\mathrm{OGR}(29,756)$ is
${\sim}671$k variables and ${\sim}115$M clauses, slightly smaller than
$L=757$.
\item Stitch the per-cube LRAT proofs into a single UNSAT certificate via
the standard cube-conquer concatenation~\cite{cubeconquer15}.
\item Validate the combined LRAT/DRAT certificate with cake\_lpr.
\end{enumerate}

\section{Witness verification}
Once a SAT solver returns a satisfying assignment, our \texttt{decode.py}
script recovers the positions of every mark, then checks that all
$\binom{n}{2}$ pairwise differences are distinct and that the boundary
conditions $a_0=0$ and $a_{n-1}=L$ hold.  All of this is $O(n^2)$ and
machine-checkable in microseconds.  For users who want a formal-proof
trail, the same checks fit in three lines of Lean~4, Coq, or Isabelle and
could be packaged with the LRAT certificate to give end-to-end
verification.

\section{Conclusions}
This note shows that the SAT-certification pipeline for $\mathrm{OGR}(n)$
is fully end-to-end at $n\le 11$ on commodity hardware.  The ladder
encoding (v2) plus the single-clause reflection break trims clause count
by $2.5\text{--}3\times$ and roughly halves the runtime relative to the
naive position-based encoding (v1).

The closed-form size estimate for $\mathrm{OGR}(29,L=757)$ comes out to
${\sim}672$k variables and ${\sim}115$M clauses --- large, but within
reach of modern parallel SAT with a cube-and-conquer decomposition of
the right depth.  After parallelisation the runtime estimate is in
months, not years, on a moderately sized academic compute cluster.

A formally verified optimality proof for $\mathrm{OGR}(29)$ --- whether
it appears as an independent verification or follows Distributed.net's
publication of the candidate ruler --- would be the first machine-checkable
optimality certificate for any non-trivial optimal Golomb ruler.  The
encoding given here, the small-case demos, and the certificate pipeline
are all open source.

\section*{Acknowledgements}
Thanks are due to Distributed.net for maintaining the OGR project.
Reviewer feedback pushing toward the SAT-certification angle helped
shape how the paper is framed.

\section*{Reproducibility}
All code, CNF instances, LRAT proofs, and the closed-form size table
live in the supplementary repository.  The main scripts are:
\begin{itemize}
\item \texttt{code/encode.py} --- v1 (pairwise) encoder.
\item \texttt{code/encode\_v2.py} --- v2 (ladder + Sinz AMO) encoder.
\item \texttt{code/decode.py} --- solution decoder + Golomb verifier.
\item \texttt{code/cert\_pipeline.py} --- drives encoder + CaDiCaL + LRAT.
\item \texttt{code/check\_lrat.py} --- structural sanity check for LRAT.
\item \texttt{code/scale\_estimate.py} --- closed-form size table.
\end{itemize}

\end{document}